\documentstyle[aps,prl,multicol,epsf,epsfig]{revtex}

\begin{document}

\newcommand{\tbox}[1]{\mbox{\tiny #1}}
\newcommand{\half}{\mbox{\small $\frac{1}{2}$}}
\newcommand{\sfrac}[1]{\mbox{\small $\frac{1}{#1}$}}
\newcommand{\mbf}[1]{{\mathbf #1}}
\newcommand{\bm}[1]{\mbox{\bf #1}}
\newcommand{\mV}{{\mathsf{V}}}
\newcommand{\mL}{{\mathsf{L}}}
\newcommand{\lB}{\lambda_{\tbox{B}}}  
\newcommand{\ofr}{{(\mbf{r})}}       

\title{Deformations and dilations of chaotic billiards,
dissipation rate, \\  
and quasi-orthogonality of the boundary wavefunctions}

\author{Alex Barnett, Doron Cohen and Eric J. Heller}

\address{Department of Physics, Harvard University, Cambridge, MA 02138}

\date{March 2000}

\maketitle


\begin{abstract}
We consider
chaotic
billiards in $d$ dimensions,
and study the matrix elements $M_{nm}$
corresponding to general deformations
of the boundary.
We analyze the dependence of $|M_{nm}|^2$ 
on $\omega=(E_n{-}E_m)/\hbar$ using semiclassical 
considerations.
This relates to an estimate of the energy dissipation rate
when the deformation is periodic at frequency $\omega$.
We show that for dilations and translations of the boundary,  
$|M_{nm}|^2$ vanishes
like $\omega^4$ as $\omega{\rightarrow}0$,
for rotations like $\omega^2$, 
whereas for generic deformations it goes to a constant.
Such special cases lead to quasi-orthogonality
of the
eigenstates on the boundary.
\end{abstract}

\begin{multicols}{2}

Chaotic cavities (billiards) in $d$ dimensions 
are prototype systems for the study 
of classical chaos and its fingerprints  
on the properties of the 
quantum-mechanical eigenstates.
As the properties of static billiards are beginning to be
understood, questions naturally arise about deformations and their time
dependence.
It is perhaps not widely appreciated that
certain deformations are very special, and that there is
a close connection between the quantum and classical mechanics of such
deformations in the case of ergodic systems.
In this paper, which
takes a fresh approach to these issues, 
we explore a special class of deformations which do not `heat'
in the limit of small frequencies.
We also establish a rather surprising relationship to a very 
successful numerical technique for finding billiard eigenfunctions.


We start with the one-particle Hamiltonian
${\cal H}_0(\mbf{r},\mbf{p}) = \mbf{p}^2/2m + V\ofr$,
where $m$ is the particle mass, $\mbf{r}$ is
the position of the particle inside the 
cavity and $\mbf{p}$ is
the conjugate momentum.
We will take the limit $V(\mbf{r}) \rightarrow \infty$ outside the
cavity, zero otherwise, corresponding to Dirichlet boundary conditions.
In this limit, the Hamiltonian is completely 
defined by the boundary shape.
The volume of the cavity we call
$\mV \equiv \mL^d$. 
Upon quantization a second length scale 
$\lB \equiv 2\pi/k$ appears,
where $k$ is the wavenumber.
For simple geometries
the typical time between collisions 
with the walls is $\tau_{\tbox{col}} \sim \mL / v$, 
where $v$ is the particle speed.
The energy is $E=\half mv^2$.
Upon quantization the eigenenergies 
are $E_n=(\hbar k_n)^2/2m$.

\begin{figure}
\centerline{\epsfig{figure=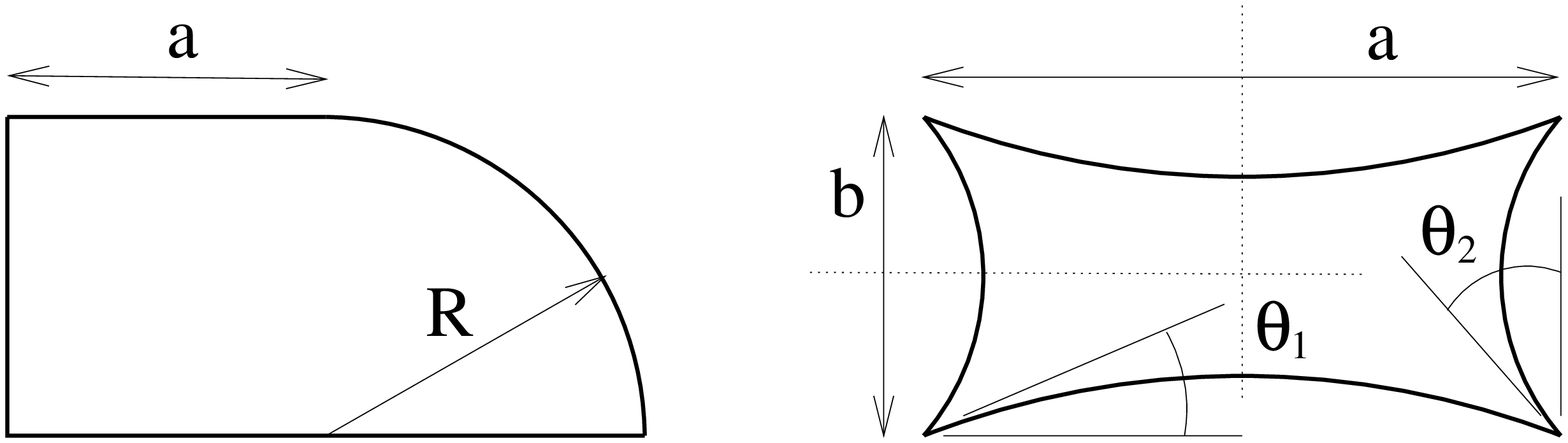,width=3.2in}}
\vspace{0.1in}
\noindent
{\footnotesize {\bf FIG. 1.}
The two-dimensional (2D) billiards under numerical study.
Left: Bunimovich quarter-stadium ($a/R = 1$).
Right: Generalized Sinai billiard ($a/b=2, \theta_1=0.2, \theta_2=0.5$).
}
\label{fig:bil}
\end{figure}

      
A powerful tool for the classical 
analysis is known as the `Poincare section'. 
Rather than following trajectories 
in the full $(\mbf{r},\mbf{p})$ phase-space, 
it is much more efficient to record only successive 
collisions with the boundary. This way 
we can deal with a canonical transformation 
(map) which is defined on a $2(d{-}1)$ dimensional 
phase space.
A similar idea is used in quantum-mechanics: 
By Green's theorem is is clear that all the information 
about an eigenstate $\psi(\mbf{r})$ is contained in the 
boundary normal derivative function
$\varphi(\mbf{s}) \equiv \mbf{n}{\cdot}\nabla\psi$,
where $\mbf{s}$ is a $(d{-}1)$ dimensional coordinate on the boundary, and
$\mbf{n}(\mbf{s})$ the outward unit normal vector.


However, unlike the classical case, the reduction 
to the boundary is not satisfactory.  
One cannot define an associated Hilbert space 
that consists of the boundary functions.
In particular, the orthogonality 
relation $\langle\psi_n|\psi_m\rangle = \delta_{nm}$ 
does not have an exact analog on the boundary.
Still, the boundary functions `live' in  
an effective Hilbert space of dimension $\sim (\mL/\lB)^{d-1}$, 
and it has been realized \cite{v+s}
that the following quasi-orthogonality
relation holds.
Define an inner product
\begin{eqnarray}   \label{eq:M}
M_{nm} \ \equiv \  
\frac{1}{2k^2}\oint \varphi_n(\mbf{s})\varphi_m(\mbf{s}) 
\ (\mbf{n}{\cdot}\hat{\mbf{D}}) d\mbf{s}
\end{eqnarray}
where $\mbf{D}(\mbf{s})=\mbf{r}(\mbf{s})$ is the
displacement field corresponding to dilation (about an arbitrary origin),
and $k_n \approx k_m \approx k$ \cite{note1}. 
It is well known that the normalization condition 
$\langle\psi_n|\psi_n\rangle = 1$  implies
$M_{nn} = 1$. We give a proof of this exact 
result in the Appendix. 
On the other hand the off-diagonal elements are only
approximately zero \cite{note2}.


The main purpose of this 
Letter is to study the band profile of 
the matrix $M_{nm}$ for a general displacement 
field $\mbf{D}(\mbf{s})$. 
In particular we want to understand why 
for special choices of $\mbf{D}(\mbf{s})$, 
notably dilations, we have quasi-orthogonality.
Later we will explain that $M_{nm}$ 
can be interpreted as the matrix element 
of a perturbation $\delta {\cal H}$
associated with a deformation of the 
boundary, such that $(\mbf{n}{\cdot}\mbf{D})\delta x$ 
is the normal displacement of a wall element,
given a control parameter $\delta x$.
In the following two paragraphs we 
explain the main motivations for our study.


The matrix elements $|M_{nm}|^2$ determine 
the rate of irreversible energy absorption by the particle
({\it i.e.}~dissipation) due to external driving.
Here `external driving' means time-dependent 
deformation of the boundary. Having exceptionally  
small $|M_{nm}|^2$ for special choices
of $\mbf{D}(\mbf{s})$, such as dilations,
translations and rotations, implies exceptionally 
small dissipation rate (`non-heating' effect). 
This observation goes against the naive kinetic picture 
that the rate of heating should not
depend on how we `shake' the boundary.
The special nature of translations and rotations 
for $\omega=0$ has been recognized in the context 
of nuclear dissipation \cite{wall,koonin}.  
Our present approach allows us to analyze the 
non-heating effect present for dilations as well,
and provide the form of the low-frequency response 
of the system in all three cases 
(dilations, translations and rotations).


There is another good motivation to study this issue. 
Recently, a powerful technique for finding clusters 
of billiard eigenstates and eigenenergies
has been found by Vergini and Saraceno \cite{v+s,verginithesis}, with a
speed typically $\sim 10^3$ greater than previous methods.
This efficiency relies on the above quasi-orthogonality relation,
the associated numerical error being
given by the deviation of $M_{nm}$
from $\delta_{nm}$.
Those authors tried to establish quasi-orthogonality   
using the identity $M_{nm} = \delta_{nm} +
[(k_m^2 - k_n^2)/2k^2] B_{nm}$, with
$B_{nm} \equiv \langle\psi_n| \, \mbf{r}\,{\cdot}\nabla \, |\psi_m\rangle$, 
and by assuming \cite{note3} that $|B_{nm}| \sim O(1)$.  
However, a naive random wave argument
would predict $|B_{nm}| \sim O(\mL/\lB)^{(d-1)/2}$.

Fig.~2 displays the band profile $|M_{nm}|^2$
for three choices of the displacement field $\mbf{D}(\mbf{s})$. 
The band profile can be regarded as either a function of
$\kappa = k_n{-}k_m$, or equivalently of
$\omega=(E_n{-}E_m)/\hbar$, related via $\omega = v \kappa$.
The three band profiles differ in their peak structure,
but also in their $\omega\rightarrow 0$ limits:
notably
for dilations $|M_{nm}|^2$ vanishes
in this limit.
Our aim is to 
understand the overall $\omega$
dependence, and the small $\omega$ behavior  
in particular.
%
%
%
%
For the calculation of band profile we
used all 451 eigenstates of the 2D
quarter-stadium (see Fig.~1) 
lying between $398 < k < 402$,
found using the method
of Vergini and Saraceno \cite{v+s}.
For this particular chaotic shape a remarkably good
basis set (size of order $\mL/\lB$)
of real and evanescent plane waves 
has been devised \cite{verginithesis},
which allows the {\em tension} error
(defined as the boundary integral of $\psi^2$)
to be typically $3 \times 10^{-11}$ in our calculation,
(maximum $2 \times 10^{-10}$ for any state).
The resulting errors in $\varphi$
manifest themselves only when
$|M_{nm}|^2$ reaches its lowest reliable
value $\sim 10^{-10}$, visible as bottoming-out
in the leftmost point of the inset of Fig.~2.


In order to understand the quantum-mechanical band profile,
we can first assume that the eigenstates look like 
uncorrelated random waves.
A lengthy but straightforward calculation \cite{frc} 
leads to the result 
\begin{eqnarray}
\label{eq:rwave} 
|M_{nm}|^2  \ \approx \   
\frac{2\langle|\cos(\theta)|^3\rangle}{\Omega_d} \ 
\frac{\lambda_{\tbox{B}}^{d{-}1}}{\mathsf{V}^2}
\oint (\mbf{n}{\cdot}\mbf{D})^2 d\mbf{s} ,
\end{eqnarray}
where the geometric factors for $d=2,3, \cdots$ are
$\Omega_d = 2\pi,4\pi \cdots$ and 
$\langle|\cos(\theta)|^3\rangle = 4/(3\pi), 1/4, \cdots$. 
If the displacement field is normalized such that
$|\mbf{D}| \sim \mathsf{L}$, then we get 
$|M_{nm}|^2 \sim (\lambda_{\tbox{B}}/\mathsf{L})^{d{-}1}$.
Note that the above result implies
that $|M_{nm}|^2$ is independent
of $\omega$.

\begin{figure}[t]
\epsfig{figure=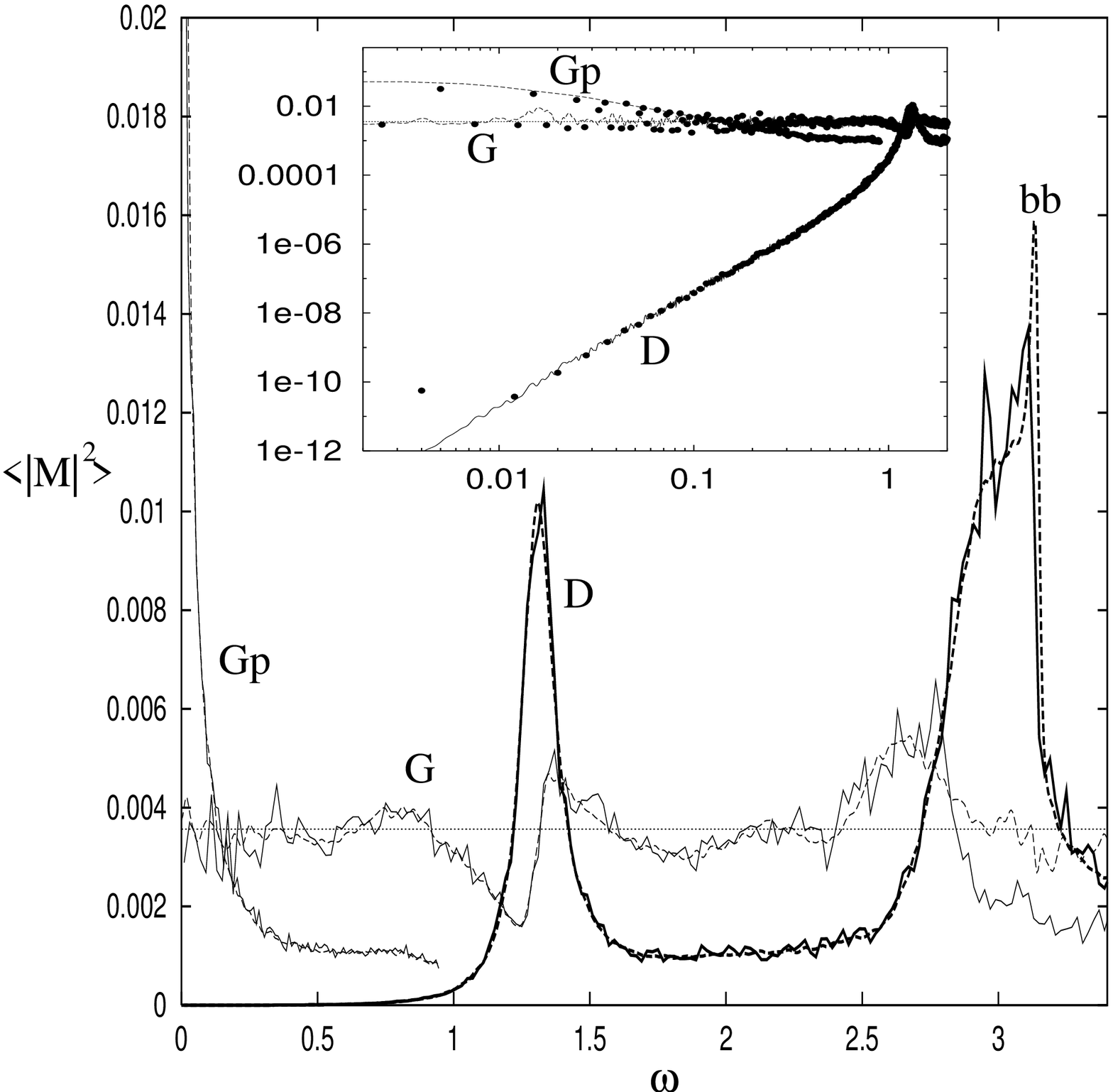,width=3.3in}
{\footnotesize {\bf FIG. 2.}
The band profile in the 2D quarter-stadium
at $k \approx 400$ for three choices of deformation field:
dilation (D), a
generic deformation (G), and
a generic deformation restricted to parallel displacement of the
stadium upper edge (Gp).
G and Gp are chosen to be volume-preserving.
In each case, the solid line is
the average $|M_{nm}|^2$ (estimation error 10\%)
versus $\omega = v( k_n-k_m)$,
with $v \mapsto 1$,
and the dashed line is the semiclassical estimate (Eq.\ref{e_11})
(estimation error 3\%).
We normalized G and Gp so that they share the 
same random wave estimate (Eq.\ref{eq:rwave}) as D; this is shown as a
horizontal dotted line.
The system-specific peak due to `bouncing-ball' orbits is labelled (bb).
The inset is a log-log plot with average $|M_{nm}|^2$ shown as points.
} 
\end{figure}


To go beyond the random-wave estimate (\ref{eq:rwave}),
we adopt a more physically 
appealing point of view. 
We include a parametric deformation of the billiard shape
via the Hamiltonian
${\cal H}(\mbf{r},\mbf{p};x) = 
\mbf{p}^2/2m + V\bm{(}\mbf{r} - x\mbf{D}\ofr\bm{)}$,
where $x$ controls the deformation.
Note that the displacement field $\mbf{D}$ is regarded 
as a function of $\mbf{r}$.
The normal displacement of a wall element is
$(\mbf{n}{\cdot}\mbf{D}) x$.
The position of a particle in the vicinity of a wall element  
is conveniently described by $Q=(\mbf{s},z)$,  
where $\mbf{s}$ is a surface coordinate 
and $z$ is a perpendicular `radial' coordinate.
We set $V\ofr = V_0$ outside the undeformed billiard;
later we take the limit $V_0 \rightarrow \infty$.
We have
${\partial {\cal H}}/{\partial x} =  
- [\mbf{n}(\mbf{s}) {\cdot} \hat{\mbf{D}}(\mbf{s})] \ 
V_0 \delta(z)$.
The logarithmic derivative with respect to $z$ of an eigenfunction on 
the boundary is $\varphi(\mbf{s})/\psi(\mbf{s})$.
For $z>0$ the wavefunction $\psi\ofr$ is a decaying
exponential. Hence the logarithmic derivative of 
the wavefunction on the boundary should be equal 
to $-\sqrt{2mV_0}/\hbar$.  Consequently one obtains 
$({\partial {\cal H}}/{\partial x})_{nm} = 
-[(\hbar k)^2/m] M_{nm}$,
Thus the band profile of $M_{nm}$ is equal (up to 
a factor) to the band profile of the perturbation 
$\delta {\cal H}$ due to a deformation of the boundary. 
See also \cite{b+w,koonin,frc}.

   
We can now use semiclassical considerations \cite{mario}. 
The application to the cavity example has been introduced 
in \cite{frc}. Here we summarize the recipe. 
First one should generate a very long (ergodic) classical trajectory, 
and define for it the fluctuating quantity
${\cal F}(t) = -{\partial {\cal H}(\mbf{r},\mbf{p};x)}/{\partial x} |_{x=0}$,
where the time-dependence of ${\cal F}$ is due to the
trajectory $\bm{(}\mbf{r}(t),\mbf{p}(t)\bm{)}$. Hence 
\begin{eqnarray} \label{e_10} 
{\cal F}(t) \ = \   
\sum_{\tbox{col}} 2mv \ \cos(\theta_{\tbox{col}}) 
\ D_{\tbox{col}} \ \delta(t-t_{\tbox{col}}) 
\end{eqnarray}
where $t_{\tbox{col}}$ is the time of a collision, 
$D_{\tbox{col}}$ stands for $\mbf{n}{\cdot}\mbf{D}$ 
at the point of the collision, and 
$v\cos(\theta_{\tbox{col}})$  
is the normal component of the particle's collision velocity. 
If the deformation is volume-preserving then 
$\langle {\cal F}(t)\rangle =0$, otherwise it is 
convenient to subtract the (constant) average value. 
Now one can calculate the correlation function 
$C(\tau)$ of the fluctuating quantity ${\cal F}(t)$, 
and its Fourier transform
$\tilde{C}(\omega) \equiv \int C(\tau) \exp(i \omega \tau) d\tau$. 
The semiclassical estimate for the matrix element is 
\begin{eqnarray} \label{e_11} 
\left\langle\left|\left(\frac{\partial
	{\cal H}}{\partial x}\right)_{nm}\right|^2 \right\rangle
\ \ \approx \ \ 
\frac{\Delta}{2\pi\hbar} \ 
\tilde{C}\left(\frac{E_n{-}E_m}{\hbar}\right)
\end{eqnarray}
where $\Delta$ is the mean level spacing.
In practice it is convenient, without loss of 
generality, to work with units such that 
in (\ref{e_10}) the time $t$ is measured in units 
of length, and we make the replacements 
$m\mapsto 1$ and $v \mapsto 1$. Then (\ref{e_11}) 
can be cast into the form 
$\langle|M_{nm}|^2\rangle \approx (\Delta_k/2\pi)\tilde{C}(\kappa)$  
where $\Delta_k$ is the mean level spacing in $k$.

Fig.~2 shows the excellent agreement between the
actual band profile and that predicted by Eq.(\ref{e_11})   
for generic deformations and dilation.
Note that there were no fitted parameters in this match. 
In all estimations of $\tilde{C}(\omega)$
we have used single trajectories of $\sim 10^6$ consecutive collisions.

Understanding the band profile of $|M_{nm}|^2$ has now
been reduced to a matter of finding
a classical theory for $\tilde{C}(\omega)$.
If we assume that Eq.(\ref{e_10})
is a train of uncorrelated impulses, then its power spectrum would be
that of white noise, namely 
$\tilde{C}(\omega) \approx \mbox{const}$.
A straightforward calculation \cite{frc} then leads
to the random wave result (\ref{eq:rwave}) already
presented.
However, in reality there are correlations in 
this train, and therefore we should expect $\tilde{C}(\omega)$ 
to have some structure on the frequency scale  
$\omega \sim 1/\tau_{\tbox{col}}$. 
Looking at Fig.~2 we see that the white noise expectation
is reasonably satisfied for one of the
`generic' deformations (G),
but not in the other two cases (D, Gp).
We also see non-universal peaks at $\omega \sim 1/\tau_{\tbox{col}} \sim 1$.
We now explain that for $\omega \ll 1/\tau_{\tbox{col}}$   
there is total failure of the white noise result  
for dilations, as well as for translations and rotations,
and discuss further complications that may arise if the 
billiard system is not strongly chaotic.

\begin{figure}
\centerline{\epsfig{figure=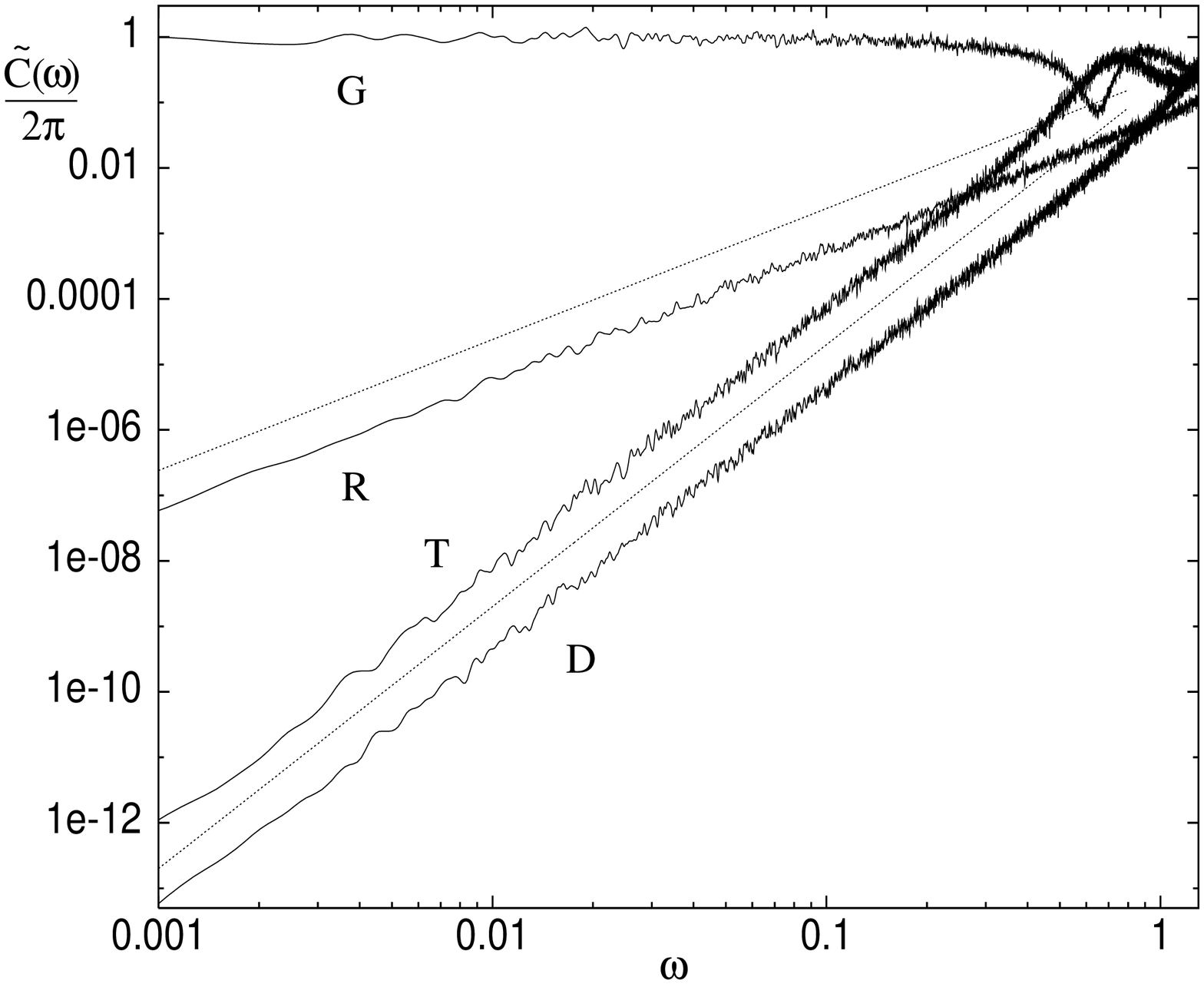,width=3.2in}}
\noindent
{\footnotesize {\bf FIG. 3.}
The classical power spectrum $\tilde{C}(\omega)$
for ${\cal F}(t)$ corresponding to a generic deformation (G),
dilation (D), translation (T), and rotation (R),
in the case of the generalized Sinai billiard with $m=v=1$.
Estimation error is 13\% for G and R, 20\% for D and T.
The two dotted lines show $\omega^2$ 
and $\omega^4$ frequency dependence, for purposes of comparison.
} 
\end{figure}

In Fig.~3 we display $\tilde{C}(\omega)$ for a different 
billiard shape, a generalized Sinai billiard (Fig.~1),
chosen because it does not suffer from the non-generic 
marginally stable orbits
found in the quarter-stadium.
Here we see very convincing evidence that for small frequencies   
we have $\tilde{C}(\omega)\approx \mbox{const}$ for 
generic deformation, while 
$\tilde{C}(\omega)\propto\omega^4$ for dilation and translation
and $\tilde{C}(\omega)\propto\omega^2$ for rotation.  
Thus the white noise expectation is indeed satisfied 
in the $\omega \ll 1/\tau_{\tbox{col}}$ regime for generic deformations,
but fails for dilations, translations and rotations,
for which $\tilde{C}(\omega)\rightarrow0$ as
$\omega\rightarrow0$.  This property is known 
(in the context of eigenvalue spectra) as `rigidity' \cite{note4}.
It implies that the train of impulses is strongly correlated,  
a result which at first sight seems inconsistent with
the assumption of chaotic motion. 
We will explain that there is no inconsistency here.


The quantity 
${\cal F}(t)=-{\partial {\cal H}}/{\partial x}$ 
is related to 
$\dot{\mbf{p}}= 
-{\partial {\cal H}}/{\partial \mbf{r}}
= -\nabla V$, 
the instantaneous force on the particle,
by ${\cal F}(t)=-\mbf{D}(\mbf{r})\cdot\dot{\mbf{p}}$.
For translations 
we have $\mbf{D}=\vec{\mbf{e}}$, 
where $\vec{\mbf{e}}$ is a constant vector  
that defines a direction in space. We can write 
${\cal F}(t)=(d/dt)^2 {\cal G}(t)$ 
where ${\cal G}(t)=-m\vec{\mbf{e}}\cdot\mbf{r}$. 
A similar relation holds for dilation $\mbf{D}=\mbf{r}$ with 
${\cal G}(t)=-\half m \mbf{r}^2$. It follows that 
$\tilde{C}(\omega)=\omega^4\tilde{C}_G(\omega)$, 
where $\tilde{C}_G(\omega)$ is the power spectrum 
of ${\cal G}(t)$. 
Assuming that ${\cal G}(t)$, unlike ${\cal F}(t)$, 
is a generic fluctuating quantity that looks like white 
noise, it follows that $\tilde{C}(\omega)$ is 
generically characterized by $\omega^4$ behavior 
for either translations or dilations. 
For rotations we have $\mbf{D}=\vec{\mbf{e}}\times\mbf{r}$, 
and we can write  ${\cal F}(t)=(d/dt) {\cal G}(t)$, 
where ${\cal G}(t)=-\vec{\mbf{e}}\cdot(\mbf{r}\times\mbf{p})$,
is a projection of the particle's angular momentum vector.
Consequently $\tilde{C}(\omega)=\omega^2\tilde{C}_G(\omega)$,
and we expect $\tilde{C}(\omega)$ to be 
generically characterized by $\omega^2$ behavior 
in the case of rotations.


In the previous paragraph we have assumed that 
generic fluctuating quantities such as 
$\mbf{r}^2$ and $\vec{\mbf{e}}\cdot\mbf{r}$ and 
$\vec{\mbf{e}}\cdot(\mbf{r}\times\mbf{p})$, as well as 
${\cal F}(t)$ for any generic deformations,
have a white noise power spectrum as $\omega \rightarrow 0$.
Obviously, this `white noise assumption' should be 
verified for any particular example. 
If the motion is {\em not} strongly chaotic, meaning that 
$C(\tau)$ decays like a power law
(say $1/\tau^{1{-}\gamma}$ 
with $0<\gamma<1$) rather than an exponential,
then the universal behavior 
is modified: we may have $\omega^{-\gamma}$ 
behavior for small frequencies.
For a generic system, for instance
the generalized Sinai billiard, we do not have this 
complication.
The stadium example on the other hand is non-generic:
the trajectory can remain in the
marginally stable `bouncing ball' orbit
(between the top and bottom edges)
for long times, with a probability
scaling as a power law in time.
Depending on the choice of $\mbf{D}\ofr$ this
{\em may} manifest itself in $C(\tau)$.
For example,
in Fig.~2 the deformation Gp     
involves a parallel displacement of the upper edge,
and the resulting sensitivity to the bouncing ball orbit
leads to large enhancement of
the fluctuations intensity $\tilde{C}(\omega{=}0)$,
and is suggestive of singular $\omega^{-\gamma}$ behavior for small $\omega$.

Finally, consider the time-dependent problem which is 
described by the Hamiltonian 
${\cal H}\bm{(}\mbf{r},\mbf{p};x(t)\bm{)}$.
It is well known that under quite 
general circumstances the dissipation 
is ohmic (${\propto}\,\,\dot{x}^2$).
See \cite{crs+vrn,frc} and references therein.
If $x(t)=A\sin(\omega t)$, linear response theory gives the
long-time heating rate
$d\langle {\cal H} \rangle / dt  =  \mu \cdot \half (\omega A)^2$. 
The dissipation coefficient $\mu$ is determined by 
the matrix elements of (\ref{e_11}), [which up to a factor 
equals $|M_{nm}|^2$], and therefore $\mu$ is 
proportional to $\tilde{C}(\omega)$.
Our results imply that $\mu$ vanishes in the 
limit $\omega\rightarrow0$ for translations.
One should not be surprised \cite{wall}, 
since this follows from Galilean invariance: 
One can view the limit $\omega\rightarrow0$ 
as corresponding to the special case of constant $\dot{x}$.
For constant nonzero $\dot{x}$ the particle(s) 
in the cavity accommodate their motion to the 
reference frame of the cavity, and there is no dissipation.
A similar argument holds for rotations.
On the other hand it is somewhat surprising that the same 
conclusion holds for dilations (the only other
shape-preserving deformation) as well.
This observation, as far as we know,
has not been introduced previously in the literature.

{\em Appendix:} 
There exist a couple of lengthy vector-identity proofs
\cite{b+w,boasman} of the normalization $M_{nn}=1$ 
for the dilation case $\mbf{D}\,{=}\,\mbf{r}$, for $d\,{=}\,2$.
Here we present a physically illuminating alternative 
that works for arbitrary $d$.
We use a phase-space-preserving definition of dilation 
operator $U(\alpha) \equiv \exp(i\alpha G/\hbar)$.  
It is generated by the hermitian operator 
$G=\half(\mbf{r}\cdot\mbf{p}+\mbf{p}\cdot\mbf{r})$.
Applying this dilation on wavefunctions gives the expansion:
\begin{eqnarray} \label{eq:dil}
U(\alpha)\psi(\mbf{r}) \approx 
\psi(\mbf{r}) + \alpha ((d/2)\psi + 
\mbf{r}\cdot\nabla\psi) + {\cal O}(\alpha^2) 
\end{eqnarray}
The operator also has the effect
$U^{\dag}\mbf{r}U = \mbox{e}^{\alpha} \mbf{r}$ 
and $U^{\dag}\mbf{p}U = \mbox{e}^{-\alpha} \mbf{p}$. 
Consider now any Hamiltonian 
${\cal H}_0 = \mbf{p}^2/(2m) + V(\mbf{r})$.
Defining the parameter-dependent version
${\cal H}(\mbf{r},\mbf{p};\alpha)= 
U(\alpha){\cal H}_0(\mbf{r},\mbf{p})U(\alpha)^{\dag}$,
it is straightforward to obtain 
\begin{eqnarray} \label{eq:dhda}
\left. \frac{\partial {\cal H}}{\partial \alpha} \right|_{\alpha{=}0} 
\ \ = \ \ \frac{\mbf{p}^2}{m} - \mbf{r}\cdot\nabla V  ,
\end{eqnarray}
whose matrix elements in the case of the 
billiard potential are
$({\partial {\cal H}}/{\partial \alpha})_{nm} = 
((\hbar k)^2/m) \ [\delta_{nm} - M_{nm}]$.
Thus the non-diagonal terms are the same as those of 
the deformation $\mbf{D}\,{=}\,\mbf{r}$.
The diagonal elements can be calculated directly 
by taking the limit $\alpha{\rightarrow}0$ of the 
expression 
$\left(\langle U\psi | {\cal H}_0 | U\psi \rangle 
- \langle \psi | {\cal H}_0 | \psi \rangle \right)/\alpha$. 
Using (\ref{eq:dil}) and the fact that
$\langle \psi | \,\mbf{r}\,{\cdot}\nabla\, |  \psi \rangle = -d/2$
one can easily show that the result 
equals zero. From here it follows that $M_{nn}=1$.


We gratefully thank Eduardo Vergini and
Mike Haggerty for important discussions.
This work was supported by
ITAMP and the National Science Foundation.

\vspace*{-0.3cm}


\end{multicols}

\begin{thebibliography}{99}
\bibitem{note1} 
As is common in the `quantum chaos' literature 
$k_n \approx k_m$ implies that $k_n{-}k_m$ is classically small,
but still $n{-}m$ may be very large.
For $n=m$, we define $k \equiv k_n$.
\bibitem{note2}
For degenerate $k_n = k_m$,
they are shown to be exactly zero in
the dilation, translation and rotation cases\cite{b+w}.
\bibitem{note3}
However, the behavior $B \sim | k_n{-}k_m |$ has since been
observed numerically
(E. Vergini, personal communication), in agreement with our results.
\end{thebibliography}
\end{document}